\documentclass{elsart3p}

\usepackage{graphicx}
\usepackage{bm}

\def\vec#1{\mathchoice{\mbox{\boldmath$\displaystyle#1$}}
{\mbox{\boldmath$\textstyle#1$}}
{\mbox{\boldmath$\scriptstyle#1$}}
{\mbox{\boldmath$\scriptscriptstyle#1$}}}
\makeatletter
\newcommand\erfc{\mathop{\operator@font erfc}\nolimits}
\def\slashchar#1{\setbox0=\hbox{$#1$}
   \dimen0=\wd0 \setbox1=\hbox{/} \dimen1=\wd1
   \ifdim\dimen0>\dimen1 \rlap{\hbox to \dimen0{\hfil/\hfil}} #1
   \else  \rlap{\hbox to \dimen1{\hfil$#1$\hfil}} / \fi}

\begin{document}

\begin{frontmatter}

\title{Pion wave function from lattice QCD vs. chiral quark models}
\thanks{Supported by the Polish Ministry of Science and Higher Education (grants
N202~034~32/0918 and N202~249235), Spanish DGI and FEDER funds (grant
FIS2008-01143/FIS), Junta de Andaluc{\'\i}a (grant FQM225-05), the EU
Integrated Infrastructure Initiative Hadron Physics Project (contract
RII3-CT-2004-506078), by the Slovenian Research Agency, by the
European RTN network FLAVIAnet (contract MRTN-CT-035482), by the Bilateral Program for Scientific and
Technological Cooperation of the Ministries of Science and Higher Education of Poland and Slovenia, and by the Slovenian-Austrian  
bilateral project under contract number BI-AT/09-10-012.}

\author[ifj,ujk]{Wojciech Broniowski},
\ead{Wojciech.Broniowski@ifj.edu.pl} 
\author[ulj,ijs]{Sasa Prelovsek},
\ead{Sasa.Prelovsek@ijs.si}
\author[ijs]{Luka \v Santelj},
\ead{Luka.Santelj@ijs.si}
\author[ugr]{Enrique Ruiz Arriola} 
\ead{earriola@ugr.es}
\address[ifj]{The H. Niewodnicza\'nski Institute of Nuclear Physics, Polish Academy of Sciences, PL-31342 Krak\'ow, Poland}
\address[ujk]{Institute of Physics, Jan Kochanowski University, PL-25406~Kielce, Poland} 
\address[ulj]{Department of Physics, University of Ljubljana, 1000 Ljubljana, Slovenia}
\address[ijs]{Jozef Stefan Institute, 1000 Ljubljana, Slovenia}
\address[ugr]{Departamento de F\'{\i}sica At\'omica, Molecular y Nuclear, Universidad de Granada, E-18071 Granada, Spain}


\date{\today}

\begin{abstract}
We analyze the equal-time Bethe-Salpeter quark wave
function of the pion obtained from a quenched lattice QCD calculation
with delocalized quark interpolators. We find 
that the result agrees remarkably well with the 
predictions of the Nambu--Jona-Lasinio model 
in all channels. We choose the quenched lattice QCD,
since it is closer to the large-$N_c$ limit of
the Nambu--Jona-Lasinio model. We also show how
transversity information, relevant for the light-cone physics, can be
obtained from our equal-time rest-frame lattice calculations.
\end{abstract}

\end{frontmatter}

\section{Introduction \label{sec:intro}}

In this paper we examine the pion quark wave functions in the Nambu--Jona-Lasinio (NJL) model
and in quenched lattice QCD, and confront both results.

Hadronic wave functions encode important information on bound states
in strong interaction physics; in particular, they provide the amplitude for a
composite hadron to have quarks in a given momentum state or,
equivalently, at a certain space-time distance.  
For systems with heavy quarks non-relativistic quantum mechanics applies and
particle number is conserved, thus much of our understanding is
directly based on wave functions. However, as a matter of principle
the wave functions cannot be directly measured experimentally and one must instead
resort to form factors, decay widths, or momentum distributions. For
light quark systems particle creation may occur, demanding a 
field-theoretic framework where further complications arise; even in the
simplest meson case, relativistic invariance requires that one uses the
conventional Bethe-Salpeter amplitude~\cite{IZ80} with fixed number of
quark field operators, a reminiscent of the approximated parton picture point of view, emphasized by the light-cone
approaches~\cite{Lepage:1980fj,Brodsky:1997de,Stefanis:1999wy}. Color
gauge invariance requires that one additionally includes link
operators~\cite{Suura:1977cw,Suura:1979ga}. For the pion, the spontaneously broken chiral symmetry is a basic dynamical
ingredient in the determination of its nonperturbative quark structure. It appears via the pertinent axial Ward-Takahashi
identities~\cite{IZ80,Pagels:1979hd} (for a review see,
e.g., \cite{Roberts:1994dr} and references therein). These important
constraints are implemented in relativistic field-theoretic chiral
quark models, such as the NJL model~\cite{Nambu:1961tp}. The regularization needs to be
carefully handled (for a relevant review see, e.g.,~\cite{RuizArriola:2002wr}). 

On the other hand, lattice QCD  solves
the bound state problem in a fundamental way. It is thus possible to make a first-principle 
non-perturbative determination but at
the expense of breaking continuum symmetries, such as the Lorentz
invariance and, quite often, chiral symmetry, due to the finite
lattice spacing. After a pioneering study~\cite{Velikson:1984qw}
the hadronic wave functions have been analyzed on the lattice on a number
of occasions~\cite{DeGrand:1987vy,Sharpe:1990rj,Chu:1990ps,Hecht:1992uq,Hecht:1992ps,Gupta:1993vp,Lacock:1994qx,Alexandrou:2002nn,vanderHeide:2003kh}. 
The axial Ward-Takahashi identities can be exactly implemented on the discrete Euclidean
lattice as shown by Ginsparg and Wilson~\cite{Ginsparg:1981bj} (see
Ref.~\cite{Gattringer:2000js,Gattringer:2000qu} for a recent practical
implementation; here we use the same method), enabling realistically small pion masses. 

\section{Bethe-Salpeter amplitude\label{sec:bsa}}

Since we attempt a comparison between
a lattice calculation, where only gauge invariant objects are defined,
and a quark model calculation with no explicit mention of the color gauge
symmetry, some remarks delineating the scope and meaning of such a
comparison are in order before presenting the actual calculations.

The Bethe-Salpeter vertex (or the quark-antiquark wave function) of the pion is given by
\begin{eqnarray}
\chi_q^b (p)  =- i \int d^4 x e^{-i p \cdot x}\langle 0 | T \left\{ q(x) \bar q(0) \right\} | \pi_b (q) \rangle, \label{bsa}
\end{eqnarray}
where $q(x)$ is the spinor field operator carrying flavor and color, 
$| \pi_b (q) \rangle$ is the pion state with isospin
$b$ and on-shell four-momentum $q$, $q^2=m_\pi^2$, and $p$ denotes the momentum of the quark field after the Fourier transform. 
While chiral quark model calculations are naturally formulated in the momentum space,
the basic objects in the Euclidean lattice calculations are
correlation functions in the coordinate space defined in Sect.~\ref{sec:lattice}, which are gauge and
renormalization group invariant at all Euclidean times.

The most general form of the quark-antiquark correlator allowed by the symmetries \cite{Braun:1989iv,Krassnigg:2009zh} has the structure
\begin{eqnarray}
\langle 0 | T \left\{ q(x) \bar q(0) \right\} | \pi_b (q) \rangle &=& {\frac{1}{4}}
{\tau_b} \\ && \hspace{-3cm} \times  [ -i \gamma_5  \Psi_P +
  i \gamma^5 \slashchar{q} \Psi_A  +  \gamma^5 \sigma^{\mu \nu} q_\mu x_\nu \Psi_T ], \nonumber
\label{eq:coor-x}
\end{eqnarray} 
where the wave functions $\Psi_a$, $a=P,A,T$, depend on the Lorentz-invariant
variables $x^2, x \cdot q $, and $q^2=m_\pi^2 $. The LHS is the inverse 
Fourier transform of the vertex
function in the Bethe-Salpeter amplitude (\ref{bsa}), which is finite and undergoes
$x$-independent multiplicative renormalization. Thus, the ratios
$\Psi(x)/\Psi(0)$ become cut-off independent as the cut-off is removed,
which on the lattice means the lattice spacing $a \to 0$. In other words, we are studying the 
renormalization-group-invariant object.

The definition (\ref{bsa}) is satisfactory for chiral quark models. In QCD, however, it
is appropriate only in the fixed point Fock-Schwinger gauge,
$x^\mu A^a_\mu(x)=0$ (the special case being the light-cone gauge $n^\mu A^a_\mu(x)=0$), where the standard 
derivatives, $\partial^\mu$, and
the covariant derivatives, $D^\mu = \partial^\mu + i g A^\mu$,
coincide. On the lattice, fixing the gauge has the problem of the Gribov
copies, as there exists no complete gauge fixing for nonabelian nonperturbative theories. Moreover, 
Elizur's theorem prevents non-vanishing vacuum expectation values of gauge variant
operators. Non-gauge-invariant bilinear operators are made gauge invariant by joining
them with a link operator, however path dependence sets
in.\footnote{See, e.g., Ref.~\cite{Broniowski:1999bz} for an illustration
  within nonlocal chiral quark models. The issue reflects the
  standard operator ordering ambiguity between $x^\mu$ and
  $p^\mu$. Only if there is a gauge fixing where the link operator
  becomes the identity, the path-independence is guaranteed.}
Furthermore, gluons carry momentum in the
pion and any gauge invariant but path-dependent definition will yield different
results (see Ref.~\cite{Negele:2000uk} for a discussion of various
possibilities). 
For definiteness, we choose the straight line prescription for the path. We also
undertake a smearing procedure of the link, as described in Sect.~\ref{sec:lattice}.

Another important issue concerns the comparison with either the {\em quenched} or
{\em dynamical} results, with the full inclusion of the fermion determinant. The
NJL Lagrangian models to the one-quark-loop level correspond to a large-$N_c$ approximation. On
the other hand, at large $N_c$ the fermion determinant is suppressed in
QCD, explaining why mesons are stable in that
limit~\cite{'tHooft:1973jz,Witten:1979kh}. The quenched approximation
contains all the leading-$N_c$ and a piece of the subleading in $N_c$
contributions, which is actually suppressed for heavy quarks. That means
that pion loops are $1/N_c$-suppressed, although not all of the $1/N_c$
contributions come from pion loops~\cite{Cohen:1992kk}. Moreover,
besides including the fermion determinant, one should also consider
higher Fock states $(\bar q q)^2$, etc., in the wave function when comparing to 
the dynamical lattice results.

In the present paper we consider the quenched lattice results and
restrict to the NJL model, as the simplest prototype of a chiral quark
model. Our study can be viewed as a useful quantitative test
of pion wave functions which can be carried out for other chiral quark models at any level of
sophistication. We recall here an early comparison of the instanton-model
hadronic wave functions to lattice results~\cite{Schafer:1994nw}.

Taking the charged pion for definiteness of the notation, 
multiplying (\ref{eq:coor-x}) with appropriate  Dirac and isospin matrices, and taking the traces
yields the relations directly used in our evaluation:
\begin{eqnarray}
\label{eq:coor-x-1}
&&\langle 0 | \bar d(0) i\gamma_5 u(x) | \pi^+ (q) \rangle =                 \sqrt{2}  \Psi_P, \\
&&\langle 0 | \bar d(0) i\gamma_5 \gamma^\mu  u(x) | \pi^+ (q) \rangle  = \sqrt{2}  q^\mu \Psi_A, \nonumber\\
&&\langle 0 | \bar d(0) \gamma_5\sigma^{\mu\nu}  u(x) | \pi^+ (q) \rangle = \sqrt{2} (q^\mu x^\nu - q^\nu x^\mu) \Psi_T. \nonumber
\end{eqnarray}
Our choice of the kinematics is $q=(m_\pi,\vec 0)$ and $x=(0,\vec r)$. In order to extract the instant-form wave functions 
$\Psi_a$ from the lattice we will consider in Sect.~\ref{sec:lattice} the matrix elements
\begin{equation}
\langle 0 | \bar d(0) \Gamma_a u(0,\vec r) | \pi^+ (m_\pi,\vec 0) \rangle = \sqrt{2}  \Psi_a(r), \label{explicit}
\end{equation}
with the explicit form of the vertices $\Gamma_P\!=\!i\gamma_5$, $\Gamma_A\!=\!i \gamma_5 \gamma_0/m_\pi$ 
and $\Gamma_T\!=\!\gamma_5\sigma_{0i} r_i/(m_\pi r^2)$.

\section{Wave functions from the NJL model}

A convenient way to determine the pion wave function
in chiral quark models is by exploiting the axial Ward-Takahashi
identity (see, e.g., \cite{RuizArriola:2002wr} for the details), relating the quark propagator $S(p)$ and
the irreducible vertex function, $\Gamma_A^{\mu,a} (p+q,p)$, corresponding to the axial current.
The spontaneous breaking of the chiral symmetry generates
a constituent quark mass $M$ given by the {\em gap
equation}, yielding $S(p)= i/(\slashchar{p}-M-m)$.
The pion wave function is extracted from the pion pole of $\Gamma_A^{\mu,a}$ in the form of an unamputated
vertex function~\cite{RuizArriola:2002wr},
\begin{eqnarray} 
\chi_q^b (p) = { i \over \slashchar{p}- M-m } g_{\pi qq} \gamma_5 \tau_b  { i \over \slashchar{p}-\slashchar{q} - M-m}.
\end{eqnarray} 
The pion-quark coupling constant, $g_{\pi qq}$, satisfies the Goldberger-Treiman relation, $g_{\pi qq}=M/f_\pi$, with $f_\pi$ denoting the pion 
decay constant. 

The approach is non-renormalizable and requires introducing a finite
cut-off which should fulfill a number of requirements concerning the gauge
and Lorentz invariance as well as causality. This is actually
crucial for many applications, including the determination of Chiral
Effective Lagrangeans~\cite{RuizArriola:1991gc,Schuren:1991sc}, Parton
Distribution Functions~\cite{Davidson:1994uv,Weigel:1999pc},
dispersion relations for two point
functions~\cite{Davidson:1995fq,Jaminon:1998de}, Parton Distribution Amplitudes and the light-cone wave
functions~\cite{Anikin:2000bn,RuizArriola:2002bp,Dorokhov:2002iu,Bzdak:2003qe}, Generalized Parton
Distributions~\cite{Theussl:2002xp,Broniowski:2007si}, Generalized Form
Factors~\cite{Broniowski:2008hx}, the photon Distribution Amplitude~\cite{Dorokhov:2006qm}, or the pion-photon 
Transition Distribution Amplitude~\cite{Lansberg:2007bu,Tiburzi:2005nj,Broniowski:2007fs,Courtoy:2007vy,Kotko:2009ij}. For practical
calculations the so-called bosonized form is more convenient and we
refer to~\cite{RuizArriola:2002wr} for further details.  

We apply the simplest twice-subtracted version of the Pauli-Villars regularization. For an
observable $A$ it amounts to the replacement $M^2 \to M^2+\Lambda^2$,
followed by the subtraction
\begin{eqnarray}
A|_{\rm reg}=A(\Lambda^2=0)-A(\Lambda^2)+\Lambda^2 \frac{dA(\Lambda^2)}{d\Lambda^2}. \label{PVpr}
\end{eqnarray}

The model has three parameters, which can be traded for $f_\pi$, $m_\pi$, and $M$.
At a fixed value of $M$ we determine $m=m_0$ and $\Lambda$ by fixing $m_\pi$ and $f_\pi$~\cite{RuizArriola:2002wr} to their
physical values $m_\pi^{\rm phys}=139$~MeV and $f_\pi=93$ MeV. In this work we use 
$M=300$~MeV, $\Lambda=790$~MeV, and $m_0=8.2$~MeV. 

The lattice simulations are performed at $m_\pi > m_\pi^{\rm phys}$, hence we need to 
increase accordingly the pion mass in the model. At not-too-large values of $m_\pi$ this may be conveniently achieved via
the Gell-Mann--Oakes--Renner relation, $m_\pi^2 \propto m$,  from where $m=m_0 (m_\pi/m_\pi^{\rm phys})^2$.
This value of $m$ is actually taken for the lattice values of $m_\pi$.

\begin{figure}[tb]
\begin{center}
\includegraphics[width=0.47\textwidth]{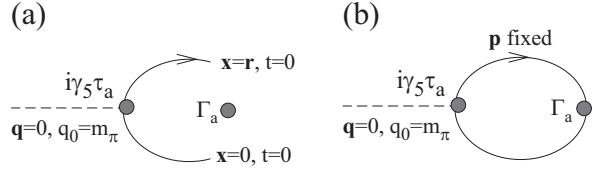} 
\end{center}
\vspace{-5mm}
\caption{Feynman diagrams for 
the one-quark-loop evaluation of the pion wave function at rest (a)~in the coordinate space and (b)~in the momentum space. 
\label{fig:loop}}
\end{figure}

Now we evaluate the quantities of our interest, $\Psi_a(r)\equiv \Psi_a(0,-r^2)$. The calculation proceeds according to the 
diagrams of Fig.~\ref{fig:loop}. Standard Feynman rules yield
\begin{eqnarray}
\Psi_a(r)= \!- \!\int \!\!\frac{d^3 p}{(2\pi)^3}e^{i\vec p\cdot 
\vec r}\!\int \!\!\frac{d p^0}{(2\pi)} {\rm Tr}[\Gamma_a S_p g_{\pi qq} \gamma_5 S_{p-q}].
\nonumber \\ \label{res1}
\end{eqnarray}
Then we perform the integration over $p_0$ and carry out the 
Fourier-Bessel transform over $\vec p$, with $p=|\vec{p}|$. The result is 
\begin{eqnarray} 
&&\Psi_P(r)\!=\!\!\int_0^\infty \!\! \frac{dp \, p^2}{2\pi^2} j_0(p r) \frac{2   N_c g_{\pi qq} \sqrt{{p}^2 \!+\!(M\!+\!m)^2}}
{p^2 +(M\!+\!m)^2-m_\pi^2/4} \biggl \vert_{\rm reg}\!\!, \nonumber \\
&&\Psi_A(r)\!=\!\int_0^\infty \!\! \frac{dp \, p^2}{2\pi^2} j_0(p r) \nonumber \\ 
&& \times \frac{N_c g_{\pi qq} (M+m)}{ \sqrt{p^2 +(M+m)^2}
{(p^2 +(M+m)^2-m_\pi^2/4)}} \biggl \vert_{\rm reg}, \nonumber \\
&&\Psi_T(r)\!=\!\int_0^\infty \!\! \frac{dp \, p^2}{2\pi^2} \frac{p}{r} j_1(p r) \label{res2} \\ 
&& \times \frac{N_c g_{\pi qq}}{ \sqrt{p^2 +(M+m)^2}
{(p^2 +(M+m)^2-m_\pi^2/4)}} \biggl \vert_{\rm reg}. \nonumber
\end{eqnarray}
The resulting ratio $\Psi_a(r)/\Psi_a(0)$, obtained numerically from Eqs.~(\ref{res2}), is plotted in Fig. \ref{fig:pwf}. 
In the chiral limit ($m_\pi=0$, $m=0$) one can carry out the integration in Eq.~(\ref{res2}) analytically, which yields 
 \begin{eqnarray}
&& \hspace{-1.5mm} \Psi_P(r)\!=\!2\Psi_T(r)\!=\!\frac{g_{\pi qq} N_c}{2 \pi^2 r} \left(-r \Lambda^2  K_0(\sqrt{\Lambda^2+M^2} r) \right . \nonumber \\ 
&&  \left . -2 \sqrt{\Lambda^2+M^2} K_1(\sqrt{\Lambda^2+M^2} r)+2 M K_1(M r)\right), \nonumber \\ 
&& \hspace{-1.5mm} \Psi_A(r)=\frac{g_{\pi qq} M N_c}{4 \pi ^2} 
\left(-\frac{\Lambda^2 r K_1\left(\sqrt{\Lambda^2+M^2} r\right)}{\sqrt{\Lambda^2+M^2}} \right . \nonumber \\
&&  \left . -2 K_0\left(\sqrt{\Lambda^2+M^2} r\right)+2 K_0(M r)\right), \label{ana}
\end{eqnarray}
where $K_0$ and $K_1$ are the modified Bessel functions.
This leads to the following asymptotic behavior at $r \to \infty$:
\begin{eqnarray}
\Psi_P(r)\sim \Psi_T(r) \sim \frac{e^{-M r}}{r^{3/2}}, \;\;\Psi_A(r) \sim \frac{e^{-M r}}{r^{1/2}}. \label{asymp}
\end{eqnarray}
We note an exponential fall-off and a longer tail in the $A$ channel than in the $P$ and $T$ channels.

One may also compute the {\em two-dimensional} Fourier-Bessel transform of Eq.~(\ref{ana}), passing from $r$ to the transverse momentum $k_T$, which 
then yields the transverse-momentum light-cone wave functions integrated over $\alpha$. We find in the chiral limit
\begin{eqnarray}
&& \hspace{-1.5mm} \Psi_P(k_T)=2 g_{\pi qq} {N_c} \\
&& \times \left(\log \left(\frac{k_T^2+\Lambda^2+M^2}{k_T^2+M^2}\right)-\frac{\Lambda^2}{k_T^2+\Lambda^2+M^2}\right) \nonumber \\
&& \hspace{-1.5mm} \Psi_A(k_T)\!=\!\frac{2 \Psi_T(k_T)}{M}\! =\! \frac{2 g_{\pi qq} \Lambda^4 M {N_c}}{\left(k_T^2+M^2\right)\!
\left(k_T^2+\Lambda^2+M^2\right)^2}. \nonumber
\end{eqnarray} 
Note that while in the coordinate representation $\Psi_P(r)\sim \Psi_T(r)$, in the $k_T$ representation $\Psi_A(k_T)\sim \Psi_T(k_T)$.

\section{Wave function from the lattice \label{sec:lattice}}

We now turn to the quenched lattice calculation of $\Psi_a({r})$, defined as (\ref{explicit})
\begin{eqnarray}
\label{j}
&&\Psi_a(r) =1/\sqrt{2}\langle 0|{\cal O}^{\vec r}_a|\pi^+(\vec q\!=\!\vec 0)\rangle, \\ 
&&{\cal O}^{\vec r}_a(\vec y,t) = \bar d(\vec x,t) {\cal P}[G] \Gamma_a u(\vec y+\vec r,t), \nonumber
\end{eqnarray}
and  illustrated in Fig. \ref{fig:loop}(a).  The interpolator annihilates quark and anti-quark at given time $t$ and at distance $\vec r$ apart. The Lorentz invariance ensures that
$\Psi_a(r)$ is independent of the arguments $\vec y$ and $t$. 
The gauge link ${\cal P}[G]$ from $\vec y$ to $\vec y+\vec r$  ensures the gauge invariance and it depends on the choice of the path, as discussed in Sect.~\ref{sec:bsa}. We choose a straight path between $\vec y$ and $\vec y+\vec r$. 
Since a straight path is unique  only  along  the lattice directions $\vec r=N a\vec e_i$, we  evaluate $\Psi_a(r)$ only for integer multiples of lattice spacing $a$, {\em i.e.}, $r=Na$. Then ${\cal P}[G]$ is a product of $N$ gauge links. For $a=P,A$ we use in fact the interpolator (\ref{j}) averaged over six lattice points at the distance $r=Na$. 
     
The correlation function is a basic object on the Euclidean lattice,
which allows one to extract the pion mass $m_\pi$, as well as its coupling to a
given interpolating composite operator $\langle 0|{\cal O}_a^r|\pi\rangle$. For our purpose we compute a correlation
function with a pseudoscalar point source ${\cal O}_P^{\vec r=0}$ at time $0$ and 
a delocalized sink with $a=P$, $A$, or $T$, ${\cal O}_a^{r}$ at the imaginary time $t=-i\tau$:
\begin{equation}
C^r_a(\tau)=\int d^3 y e^{i\vec q\cdot \vec y}\langle 0|{\cal O}^r_a(\tau ,\vec y){\cal O}^{r=0 \dagger}_P(0,\vec 0)|0\rangle \label{corr}
\end{equation}
where we project on the total momentum $\vec q=\vec 0$.

The quantities related
to the pion can be extracted after the complete set
of physical states is inserted into (\ref{corr}), yielding
\begin{eqnarray}
C^r_a(\tau) \!=\!\!
\sum_n \!\frac{\langle 0|{\cal O}_a^r(0)|n_{\vec q}\rangle\langle n_{\vec q}| 
{\cal O}_P^{r=0 \dagger}(0)|0\rangle}{2E_n(\vec q)}e^{-E_n(\vec q)\tau}. \nonumber \\
\end{eqnarray}
The ground-state pion at rest
dominates at large $\tau$,
\begin{eqnarray}
&&C^r_a(\tau)\stackrel{\tau\to\infty}{\longrightarrow} w^r_a
~e^{-m_\pi\tau}, \\
&&w^r_a=\frac{\langle 0|{\cal O}_a^r|\pi
  \rangle\langle \pi| {\cal O}_P^{r=0~\dagger}|0\rangle}{2m_\pi}. \nonumber
\end{eqnarray}
From a single exponential fit to the lattice correlation functions
at large $\tau$ we extract $m_\pi$ and $w_a^r$ for a range $r=[0,a,..,4a]$. 
 We use several current quark masses $m$, which corresponds to  $m_\pi$ 
in the range $345-740$~MeV. 

The extracted pion mass does not numerically depend on $r$ nor the choice of
the channel $P$, $A$, or $T$, as expected. The ratios $\Psi_a(r)/\Psi_a(0)$ 
for P and A channels are extracted from the identity
\begin{equation}
\frac{\Psi_a(r)}{\Psi_a(0)}=\frac{ \langle 0|{\cal O}_a^r|\pi
  \rangle}{\langle 0|{\cal O}_a^{r=0}|\pi
  \rangle}=\frac{w_a^r}{w_a^{r=0}}
\end{equation}
and plotted in Fig. \ref{fig:pwf}. The wave function 
$\Psi_T$ can not be evaluated at $r=0$ on the lattice and 
 we normalize it arbitrarily, such that the NJL model results and lattice values 
 at $r=a$ agree  ({\em cf.} Fig. \ref{fig:pwf}).

Finally, we provide some details of the lattice simulation. We use quenched lattice QCD, since it is closer to the large-$N_c$ limit of 
the NJL model, as explained in Sect.~\ref{sec:bsa}. 
We use 100 gauge configurations generated by the  L\"{u}scher-Weisz gauge 
action \cite{Luscher:1984xn}, as described in \cite{Burch:2006dg}. 

We use so called HYP smearing on the gauge configurations \cite{Knechtli:2000ku}.
This replaces a gauge link between the neighboring points on the lattice
with a ``fat'' link. The fat link is a sum of links within hypercubes attached
to the original link only. Since this smearing is local, it does not destroy short distance quantities, while it smooths out the large local fluctuations of the gauge field.

The lattice spacing $a\simeq 0.148$~fm is determined from the Sommer parameter, and the lattice volume is
$16^3\times 32$. We employ the chirally improved valence quarks
\cite{Gattringer:2000js,Gattringer:2000qu,Burch:2006dg}, which have
good chiral properties.

\section{NJL vs. quenched lattice}

\begin{figure}[tb]
\begin{center}
\includegraphics[width=0.47\textwidth]{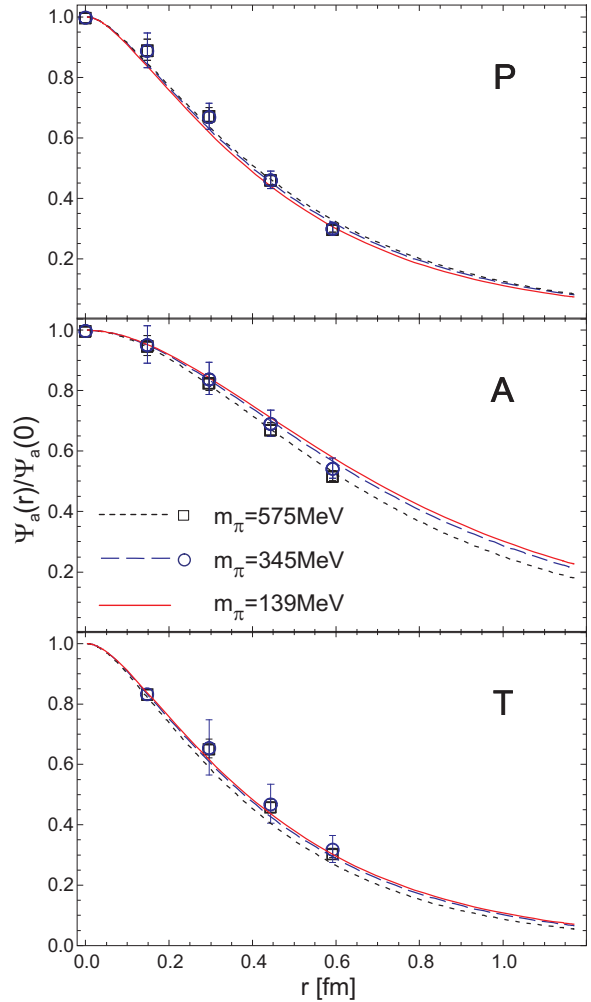} 
\end{center}
\caption{The components of the rest-frame pion wave
  function, normalized to unity at the origin, $\Psi_{a}({r})/\Psi_{a}({0})$,
  for $S$, $A$, and $T$ channels, 
  evaluated in the NJL model at $M=300$~MeV and compared to the quenched lattice 
  data. The shown points are at $m_\pi=345$ and $575$~MeV, while the model calculation includes also the 
  case of the physical pion mass.}
\label{fig:pwf}
\end{figure}

Our results are presented in Fig. \ref{fig:pwf}, with the lines showing the model and the points the lattice calculation. 
In the $P$ and $A$ channels we normalize $\Psi_{a}({r})$ to its value at the origin, while for the $T$ channel we normalize to the model 
prediction at $r=a$.
We note that the results of the NJL model agree
remarkably well with the lattice determination in all three channels. The dependence on $m_\pi$ 
is very weak. The insensitivity of the
lattice wave functions to the value of the pion mass confirms previous
findings~\cite{Chu:1990ps,Hecht:1992uq,Alexandrou:2002nn}. As expected from Eq.~(\ref{ana}), the
pseudoscalar and tensor wave functions are identical within the error bars, while the
axial wave function is found to be wider. This agrees with the 
different asymptotic behavior of Eq.~(\ref{asymp}).

We have also checked 
that the model results are almost independent of $M$ when this parameter is in
the reasonable range, $M=250-350$~MeV. 

\section{Rest-frame kinematics and transversity}

We end this paper with general remarks relating the equal-time wave functions 
presented above to the light-cone wave functions in the impact-parameter 
representation, also known as {\em transversity} wave functions. 
Formally, we can write the transformation
\begin{eqnarray}
&& \langle 0 | T \left\{ q(x) \bar q(0) \right\} | \pi_b (q) \rangle = 
 \frac{i \gamma_5 \tau_b}{4} \!\! \int_0^1 \! d\alpha e^{i (2\alpha-1) q \cdot x } \! \times \nonumber \\ 
&& [ -\Phi_P(\alpha, x^2) + \slashchar{q} \Phi_A (\alpha, x^2) - i \sigma^{\mu \nu} q_\mu x_\nu \Phi_T (\alpha,  x^2)], \nonumber \\
  \label{eq:coor-x-2}
\end{eqnarray} 
where $\alpha$ is the Feynman parameter~\cite{Braun:1989iv}. The presence of a gauge link operator between the quarks is implicit. Comparison to 
(\ref{eq:coor-x}) yields by construction  
\begin{eqnarray}
\Psi_a(x\cdot q, x^2) = \int_0^1 d\alpha e^{i (2\alpha-1) q.x}\Phi_a ( \alpha,x^2 ). 
\end{eqnarray}
As a matter of
principle, all scalars $\Psi_a$ in Eq.~(\ref{eq:coor-x-2})
depend on the scalar variables $x^2$, $q^2$, and $ x \cdot q $, hence we are free
to choose any reference frame to evaluate $\Psi_a$.
In the rest-frame, or equal-time (ET) kinematics, used in the previous sections of the paper, we take
$x=(0,\vec{r})$ and $q=(m_\pi, 0)$, whence $x^2 = - r^2$ and $ x \cdot q =0$. In the 
infinite-momentum-frame kinematics $(q_0,{\bf q})=\lim_{q_z \to \infty}(\sqrt{m_\pi^2+q_z^2}, q_z)$
and on the light cone (LC), where $x^+=0$, one has $x \cdot q = q^+ x^-$, $x^2=-r^2$. The parameter $\alpha$ acquires the 
meaning of the light-cone momentum fraction of pion carried by one of the quarks.  
By comparing the two calculations we find 
$\Psi^{\rm ET}_a(0, -r^2) = \int_0^1 d\alpha e^{i (2\alpha -1) q^+ x^-}\Phi^{\rm LC}_a (\alpha, -r^2 )$.
For the chosen kinematics $q^+ x^-=q \cdot x=0$, hence
\begin{eqnarray}
\Psi^{\rm ET}_a(0, -r^2) = \int_0^1 d\alpha \Phi^{\rm LC}_a (\alpha, -r^2 ). \label{trarel}
\end{eqnarray}
In other words, our rest frame calculation allows for a direct
determination of the transverse-coordinate (impact parameter) dependence of the light-cone wave-function
integrated over the $\alpha$ parameter.  
A similar property for the Generalized Parton Distributions
was suggested for the nucleon~\cite{Miller:2007uy} and the
pion~\cite{Broniowski:2007si}. The relation between the equal-time and light-front wave functions
has been analyzed recently \cite{Miller:2009fc} in the momentum space, where the
transversity relation (\ref{trarel}) cannot be explicitly seen.
 
It would also be useful to verify the equal time-light cone
transversity connection on the lattice. While there exist transverse
lattice calculations~\cite{Dalley:2002nj}, their focus was on the Distribution Amplitude,
$\Psi(\alpha, 0) = \varphi(\alpha)$, leaving out the transverse
dependence. Some results where also presented in~\cite{Abada:2001if}.

The transversity relation (\ref{trarel}) 
is explicitly verified for the NJL light-cone wave
function~\cite{RuizArriola:2002bp}.

\section{Conclusions}

Here are our main points:

\begin{itemize}

\item The leading-$N_c$ chiral quark model interpretation of the quenched
lattice data is not only qualitatively correct, but also remarkably
accurate. This is yet another manifestation of the fact that the spontaneously broken chiral symmetry is the key 
dynamical factor in the pion dynamics. The quality of the agreement suggests that the $1/N_c$ contributions in the 
quenched smeared lattice simulations are small.   

\item The dependence of both the lattice and model results on the value of the pion mass is very small in the broad tested range 
$m_\pi=345-740$~MeV.

\item The pseudoscalar and tensor quenched pion wave functions are equal on the lattice within the error bars. In the model they are equal in the chiral limit and nearly equal for the used values of $m_\pi$. 
 
\item The asymptotic fall-off of the pion wave functions in the model is exponential, $\sim \exp(-M r)/r^p$, with $p=3/2$ in the pseudoscalar and tensor channels, while $p=1/2$ in the axial channel. This qualitatively complies to the lattice data, where the axial wave function exhibits a longer tail. 

\item A general result, which originates from the Lorentz invariance, 
concerns the utility of the equal-time rest-frame smeared lattice simulations to determine the
transversity information relevant for the light-cone physics. The integrated 
light-cone (infinite-momentum) pion wave functions in the
impact-parameter space, $\int_0^1 d \alpha \Phi_a^{\rm LC}(\alpha,b)$, coincide with our equal-time rest-frame wave
functions evaluated at the same quark-antiquark separation, $\Psi_a^{\rm ET}(0,-r^2)\mid_{r=b}$. 

\end{itemize}

\bigskip
\bigskip

One of us (SP) thanks Bojan Golli, Christian B. Lang, and Mitja Rosina for useful discussions.  We are
grateful to the Bern-Graz-Regensburg Collaboration for providing the quenched lattice 
configurations and the quark propagators, indispensable for this work. They were
generated in the Leibniz Rechenzentrum, Garching and the ZID, Graz.

\bibliographystyle{h-elsevier}
\bibliography{smear}

\end{document}